\documentclass[10pt,letterpaper,twocolumn]{article} %% two column, final layout

\usepackage{ol2}
\usepackage[draft]{hyperref}
\usepackage{amsmath}
\usepackage{amssymb}
\usepackage{multirow}
\usepackage{color}
\usepackage{ulem}

\begin{document}

\twocolumn[ %% activate for two-column option

\OSAJNLtitle{Diffusive Suppression of AC-Stark Shifts in Atomic Magnetometers}

\OSAJNLauthor{I. A. Sulai,$^{1}$, R. Wyllie $^{1,3}$, M. Kauer $^{1}$,  G. S. Smetana $^{1}$, R. T. Wakai,$^{2}$ and T. G. Walker,$^{1,*}$}
\address{
$^1$Department of Physics, University of Wisconsin-Madison, 1150 University Avenue, Madison, WI 53706, USA\\
$^2$Department of Medical Physics. University of Wisconsin-Madison, 1111 Highland Avenue, Madison, WI 53706, USA\\
$^3$Current address: U.S. National Institute of Standards and Technology, Gaithersburg, Maryland 20899.\\
$^*$Corresponding author: tgwalker@wisc.edu
}

\begin{abstract}
In atomic magnetometers, the vector AC-Stark shift associated with circularly polarized light generates spatially varying effective magnetic fields which limit the magnetometer response and serve as sources of noise. We describe a scheme whereby optically pumping a small sub-volume of the magnetometer cell and relying on diffusion to transport polarized atoms allows a magnetometer to be operated with minimal sensitivity to the AC-Stark field.
\end{abstract}

] %% activate for two-column option

Spin Exchange Relaxation Free (SERF) atomic magnetometers (AMs) are presently the most sensitive magnetic field detectors. With demonstrated sensitivities $< 1$ fT/$\sqrt{\rm Hz}$ \cite{BudkerRomalis}, they have been used in applications such as geomagnetism \cite{Dang}, biomagnetism \cite{Wyllie,Wyllie2,Sander}, and tests of fundamental physics \cite{Brown}. SERF magnetometers reach their maximum sensitivity when the spin precession rate is much smaller than the total spin relaxation rate \cite{Allred}. This requires operating the AMs in total magnetic fields on the order of $B_m\,\sim\,10$\,nT or less.

Typically, AMs are optically pumped by  absorption of near-resonance circularly polarized  photons. However, virtual absorption of the photons also gives rise to an AC-Stark shift. As described in \cite{Appelt99,Happer72}, the vector Stark shift for circularly polarized light assumes the form $ \delta H_v = \hbar \delta \Omega_v \bf{s}\cdot\bf{S}$, where $\bf{s}$ is the photon spin and $\bf{S}$ the electron spin. $\delta H_v$ thus perturbs the levels just like a magnetic field \cite{CCT72,Happer72}. Such fields are often comparable to $B_m$ in strength and  can degrade the AM response.

High sensitivity magnetic sensing applications such as fetal magnetocardiography (fMCG) \cite{Wyllie,Wyllie2} often require a large number of gradiometer channels with $\sim$\,cm baseline . For these, the unparalleled sensitivity of AM is only exploited to the extent that background signals can be suppressed, since the magnetic background can be much larger than the signal of interest. 
%For example, in studies of fetal magnetocardiography (fMCG), the mother's heart beat is typically an order of magnitude larger magnetic signal and is most effectively removed using a spatial filter \cite{Wyllie}.
Common mode magnetic field fluctuations between different channels can be subtracted in a gradiometer. However, AC-Stark fluctuations are often uncorrelated and therefore limit the common mode rejection of the gradiometers.

Although operating the pump laser on resonance results in a vanishing AC-Stark shift, doing so while maintaining adequate atom density and polarization is difficult. The sensitivity of the AM increases with atom number, and is maximized when the pumping is such that the polarization is 1/2. A near resonant pump does not propagate through high optical density (OD) cells unless the pumping rate is so high that it reduces AM sensitivity. Tuning the pump off-resonance to optimize the sensitivity then gives rise to a substantial AC-Stark field.  A real compensating field can be added to cancel the average AC-Stark field. But, because of the Gaussian profile of the pump laser and it's absorption in the atomic vapor, the spatial distribution of the AC-Stark field is non-uniform so that AC-Stark gradients still persist.

In this paper, we demonstrate the use of diffusive atom transport to greatly suppress AC-Stark effects.  Our approach is to optically pump the atoms very strongly in a small sub-volume of the AM cell, then rely on diffusion to transport the polarized atoms to regions of the cell with little or no AC-Stark field.  We demonstrate a reduction of the AC-Stark field by more than an order of magnitude, with minor impact on the AM sensitivity.

Diffusion to the walls and collisional spin-relaxation are dominant spin relaxation mechanisms in AMs. Their relative importance can be estimated by comparing the diffusion length $z_d = \sqrt{D_0 / \Gamma}$ to the cell radius $r_0$, where $D_0$ is the diffusion constant, $\Gamma$ is the transverse spin relaxation rate, and $z_d$ is the characteristic distance traversed by polarized atoms before participating in spin destroying collisions.  When the diffusion length is comparable to the cell radius, polarized atoms can be readily transported from a localized optical pumping region to the AC-Stark-free remainder of the cell volume, allowing for sensitive magnetometry with substantially reduced effects of AC-Stark shifts. The pumping region is localized by choosing a pump laser with waist $w_p \ll z_d$. We call an AM operating in this mode a diff-SERF, and used an array of such diff-SERFs in our recent detection of fMCG \cite{Wyllie2}.

A  SERF AM can be understood from angular momentum conservation \cite{Ledbetter}. The total polarization distribution $\mathbf P(\vec{r})$ evolves as
\begin{align}
\frac{\partial \mathbf{P}(\vec{r})}{\partial t} =  &D_0 \nabla^2 \mathbf{P}(\vec{r}) + \frac{1}{q(P)} \Bigl[ R(\vec{r})(\hat z 
 - \mathbf{P}(\vec{r}))\nonumber \\ +& \mathbf{\Omega} \times \mathbf{P}(\vec{r}) + \Omega_v (\vec{r}) (\hat{z} \times \mathbf{P}(\vec{r})) - \Gamma~ \mathbf{P}(\vec{r}) \Bigr]   
\label{eqODE}
\end{align}
for optical pumping along the $\hat{z}$ direction and for the polarization defined as $\mathbf{P} =  2\mathbf{\langle S\rangle}$. Here $q(P)$, the `slowing down factor',  accounts for the storage of angular momentum in the nucleus \cite{Allred,HapperBook}.  The optical pumping term $  R(\vec{r})(\hat z - \mathbf{P}(\vec{r}))$  serves as the source of angular momentum.  $\bf{\Omega} \times \bf{P}$ describes spin precession in the external magnetic field while the $\Omega_v (\vec{r}) (\hat{z} \times \mathbf{P}(\vec{r}))$ term describes the precession in the Stark field. The $- \Gamma \bf{P}$ and $D_0 \nabla^2 \mathbf{P}$ terms describe spin relaxation due to atom\,-\,atom and atom\,-\,wall collisions respectively. 

To qualitatively understand the AM performance, we approximate the diffusion term with an effective relaxation rate $\Gamma_{\rm diff}$. Defining $\Gamma'(\vec{r}) = \Gamma +  \Gamma_{\rm diff} + R(\vec{r})$ and  $\tilde{\Omega}_z(\vec{r}) = \Omega_z + \Omega_v(\vec{r})$, we note that in the limit $|\vec{\Omega}| \ll |\Omega_v|$, the detected transverse polarization is
\begin{equation}
P_x(\vec{r}) \sim P_z(\vec{r}) \frac{\Gamma'(\vec{r}) \Omega_y + \Omega_x \tilde{\Omega}_z(\vec{r})}{\Gamma'(\vec{r})^2+ \Omega_v^2(\vec{r})}.
\label{eq:MagEq}
\end{equation}
This shows that the non-zero polarization distribution $P_z (\vec{r})$ established by optical pumping results in a AM sensitive to $\Omega_y$ with a response $\partial P_x/ \partial \Omega_y$. The denominator of Eq. \ref{eq:MagEq} implies that the response is reduced in regions where the pumping and AC-Stark precession rates are large. Furthermore, the numerator indicates that if the AC-Stark field is non-zero, the AM becomes sensitive to fields in the $\vec{x}$ direction, an often undesirable feature. AC-Stark effects thus degrade magnetometer performance by reducing the AM response as well as creating sensitivity to fields in a secondary direction.

%In the diff-SERF with spatially localized  optical pumping with a  pump waist  $w_p \ll  z_d $,  polarized atoms diffuse out of the pumping region where they remain sensitive to transverse fields   but experience no AC-Stark shift. Although atoms inside the pumping region experience large Stark shifts, the correspondingly large pumping rates lead to rapid spin dephasing and consequently diminished sensitivity to magnetic fields.
\begin{figure}[h!]
\centering
\includegraphics[width=1.0\linewidth]{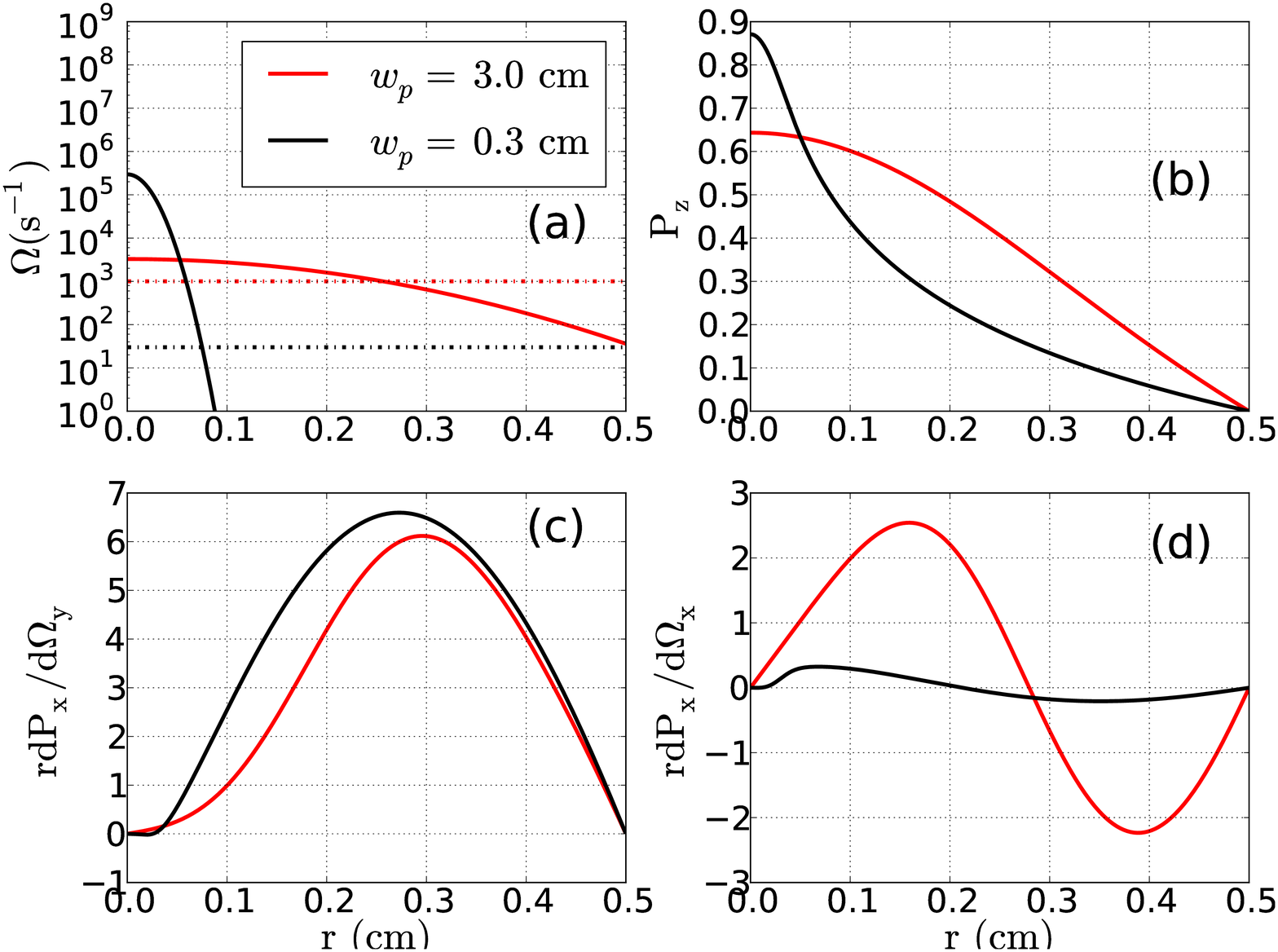}
\caption{SERF simulation (Eq.~ \ref{eqODE}) with a pump laser having power 0.2 mW, waist $w_p$ and detuning 11\,GHz. (a) The vector AC-Stark field $\Omega_z(r)$. The dotted lines show the value of the compensation field.
(b) The spatial distribution of the polarized atoms $P_z(r)$.  
(c) Radially weighted distribution of magnetometer response $ \partial P_x (r)/\partial \Omega_y$.  
(d) Radially weighted distribution of undesired sensitivity $\partial P_x(r)/\partial \Omega_x$. 
}
\label{fig:model1}
\end{figure}

Figure~\ref{fig:model1} shows a numerical solution to Eq.~\ref{eqODE} in a $1\,$cm diameter cylindrical cell and realistic magnetometer parameters,
The figure of merit for a magnetometer that detects $P_x$ is $\int{ g_{pr}(\vec{r}) \partial P_x/\partial \Omega_y ~d^3 r}$, where the probe spatial profile is $g_{pr}(\vec{r})$. The probe is chosen to have a waist $w_{pr} \gtrsim z_d$, allowing for the sampling of most of the polarized atoms.  The measure of the sensitivity to vector AC-Stark shifts is  the magnitude of the uniform compensation field $\Omega_z \hat{z}$ which is applied to the AM to null the unwanted response $\int{ g_{pr}(\vec{r}) \partial P_x/\partial \Omega_x ~d^3 r} \to 0$.  Thus $\Omega_z$ is  the spatially-averaged AC-Stark precession rate.

We show two cases in  Fig. \ref{fig:model1}:  `big pump', with $w_p\,\sim\,z_d$,  and `small pump', with $w_p\,\ll\,z_d$.  In panel (a.), we note that the spatially averaged AC-Stark precession rates are respectively $\sim\,1000\,s^{-1}$ and $\sim 25~s^{-1}$ -- indicating a $40\times$ reduction in AC-Stark  sensitivity for the small pump. This can be understood from  panels (b-c).  When using the small pump,  the atoms still diffuse and fill the volume as indicated by the steady-state polarization distribution $P_z(r)$. Furthermore,  the magnetic field response $\partial P_x/ \partial \Omega_y$ remain comparable in the `small pump' and `big pump' cases. From panel (d), we see that the 'big pump' case generates significantly more sensitivity to fields along $\vec{x}$ than the diff-SERF.

A key advantage to having reduced Stark shift sensitivity is that the noise associated with the effective field is  also reduced.  While the time and volume average of the AC-Stark shift can be compensated for, the instantaneous fluctuations in the field are not. Thus, from our model, the  diffusion mode pumping AM would be about a factor of 40 less susceptible to AC-Stark noise.

\begin{figure}[h!]
\centering
\includegraphics[width=0.75\linewidth]{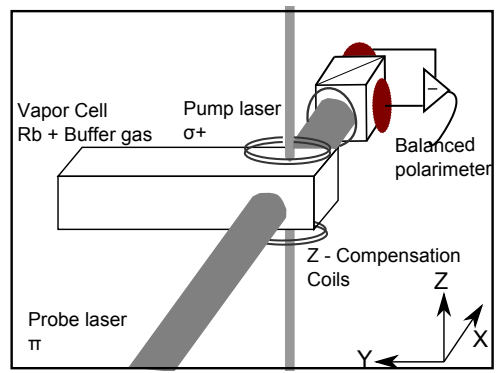}
\caption{Schematic of AM. Atoms polarized by the small but intense pump  diffuse out into the remainder of the cell to be detected by the probe in places where their spin-precession is not degraded by the pumping light. }
\label{fig:setup}
\end{figure}

To experimentally demonstrate the diff-SERF as illustrated in Fig.~\ref{fig:setup}, we used an uncoated vapor cell with $1\,$cm $\times$ $1\,$cm cross section containing $^{87}$Rb atoms in approximately 50 Torr Nitrogen + 15 Torr Helium buffer gas. The cell is pumped with $\sigma^+$ Rb - D1 light at 795\,nm along the $\hat{z}$ axis. A balanced polarimeter is used to detect $P_x$ by the Faraday rotation of linearly polarized probe light at 780\,nm, near the Rb - D2 line. The AM is operated inside a magnetically shielded room and is surrounded by three pairs of coils used to apply compensating magnetic fields in orthogonal directions \cite{Wyllie}. 

We present measurements on the diff-SERF at temperatures of  80\,$^\circ$C and 170\,$^\circ$C, corresponding to  densities of $\sim 10^{12}\,\mathrm{cm}^{-3}$ and $\sim 10^{14}\,\mathrm{cm}^{-3}$ respectively. The high optical depth in the latter case (OD$\,\sim\,200$ on resonance), accentuates the impact of pump photon absorption and the resulting longitudinal magnetic field gradients from the AC-Stark shift.

\begin{figure}[ht!]
\begin{center}
\includegraphics[width=0.9\linewidth]{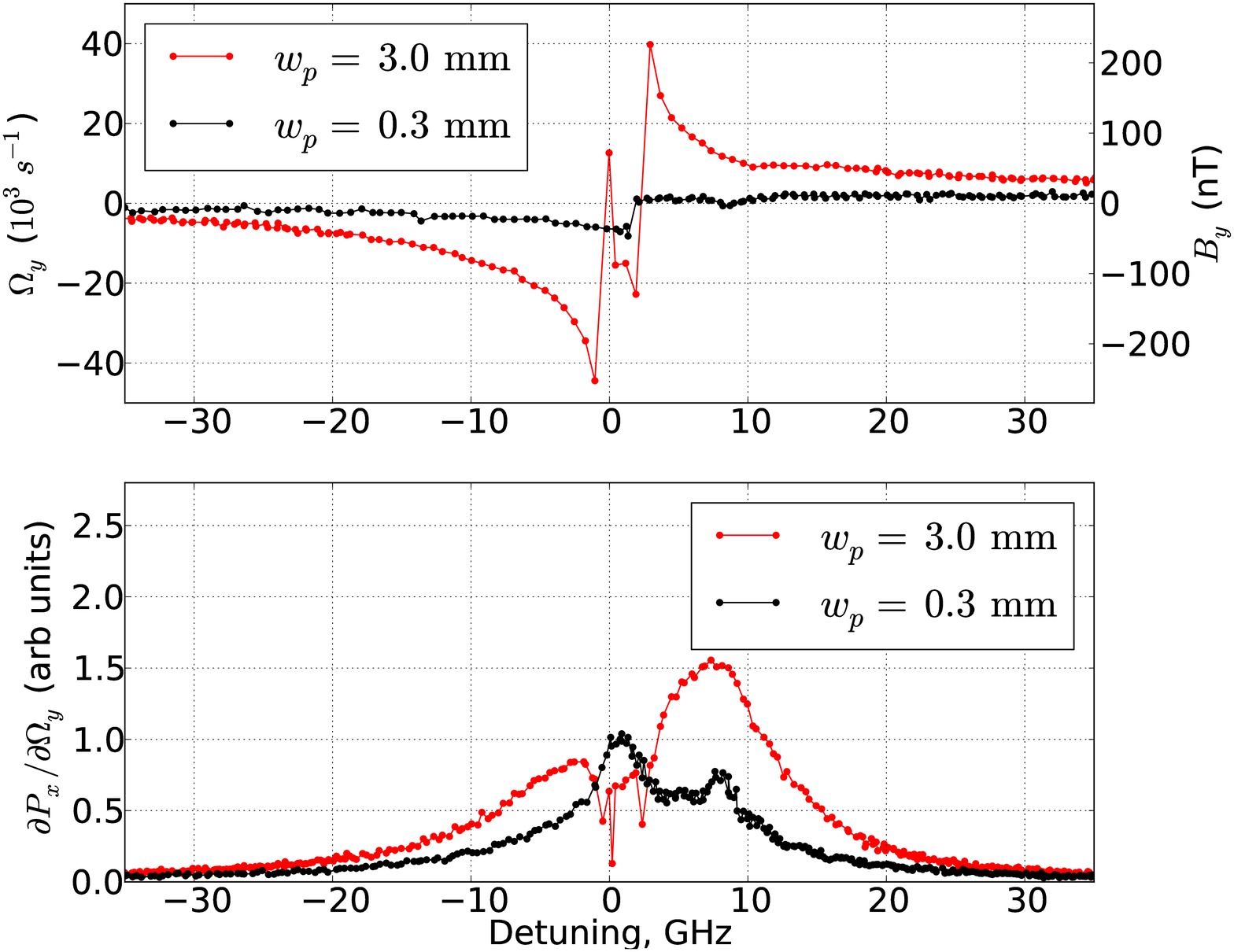}
\end{center}
\caption{ Measurements taken with T = 80 $^\circ$C, for different pump laser waists $w_p$. Top: measured average AC-Stark field 
vs. pump laser frequency.  Bottom: AM response vs. frequency. At this temperature and vapor cell composition we calculate $\Gamma = 7~\mathrm{s^{-1}}$, and $z_d = 0.5\,\mathrm{cm}.$}
\label{fig:LowT}
\end{figure}

\begin{figure}[ht!]
\begin{center}
\includegraphics[width=0.9\linewidth]{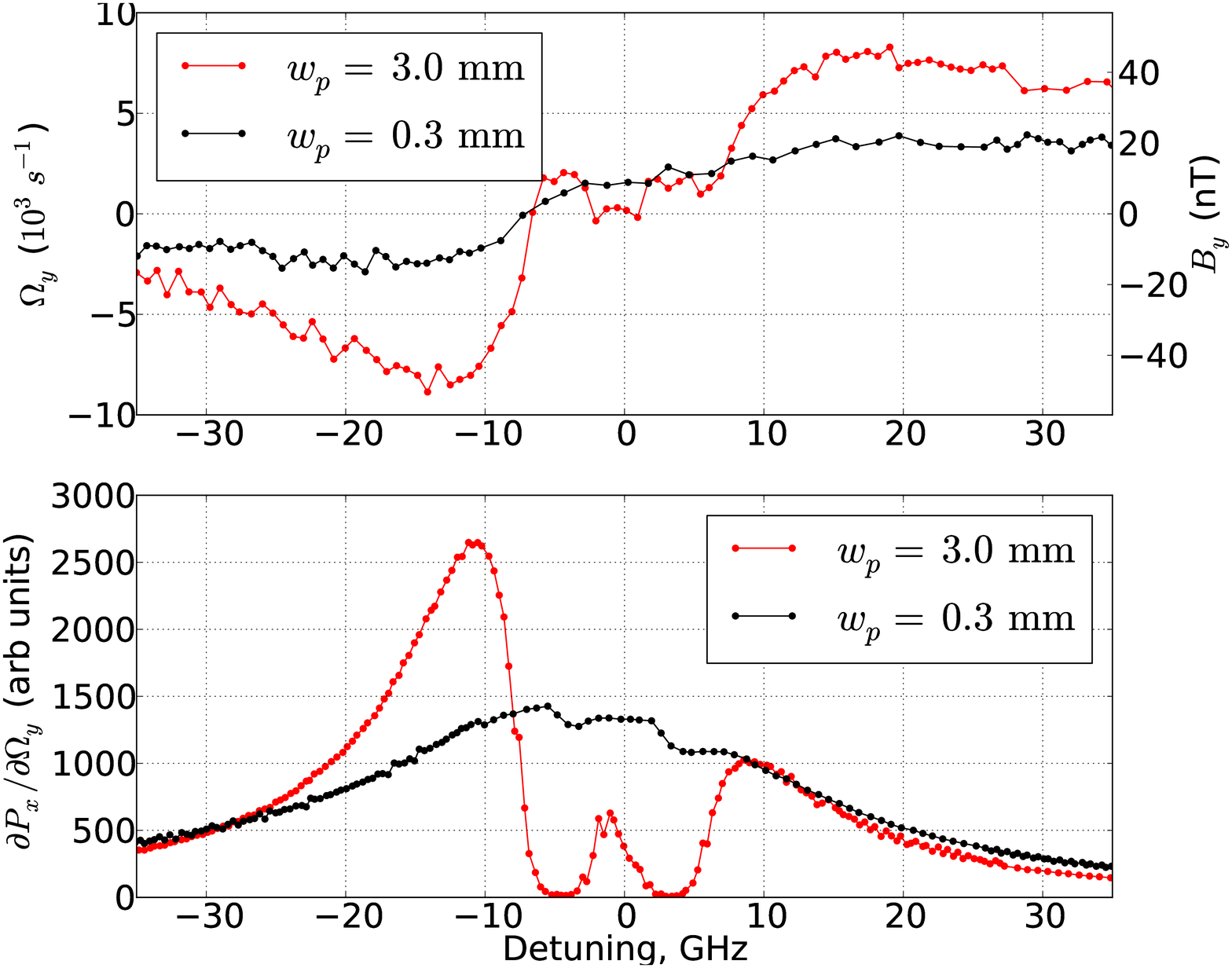}
\end{center}
\caption{Same as Fig.~\ref{fig:LowT}, but  T = 170 $^\circ$C. We calculate $\Gamma = 100~\mathrm{s^{-1}}$ and $z_d = 0.1\,\mathrm{cm}.$}
\label{fig:HighT}
\end{figure}

For each temperature and laser configuration, we  scanned the pump laser frequency %over a range 
$ \pm 30\,$GHz about the atomic resonance. At each setting, we adjusted the compensation field to cancel the average AC-Stark shift, with results  plotted in the upper panels of Fig.'s \ref{fig:LowT} and \ref{fig:HighT}. The characteristic dispersive lineshape of the vector AC-Stark Shift is clear. At each frequency setting, we also measured the response of the magnetometers $\partial P_x/ \partial \Omega_y$,  shown in the lower panels of Fig.'s \ref{fig:LowT} and \ref{fig:HighT}.

In agreement with the model discussed above,  Fig.'s \ref{fig:LowT} and \ref{fig:HighT} show that the sensitivity to AC-Stark shifts is substantially suppressed in the diff-SERF mode, while the response of the AM changes by less than a factor of 2. We also note that the dependence of the response on laser frequency detuning has significantly more structure for the large pump case. This difference is especially dramatic for $T =170 ^0 C$. For the `large pump' case, we also observe an asymmetry in the detuning dependence of the large pump. We suspect that this is due to Stark shifts from asymmetrical optical pumping, since we operate in a vapor cell with resolved hyperfine structure.

These results  show  that the diff-SERF is a viable method for evading the consequences of the vector AC-Stark shift in SERF atomic magnetometers.
The diff-SERF offers a number of important technical advantages for applications requiring high magnetic field sensitivity. The reduced vector Stark shift allows   the pump laser frequency to be optimized independently of the compensating magnetic fields. The diff-SERF pump laser works over a large frequency range with little performance loss -- an attractive feature when pumping an array of several different cells with the same laser.  In addition, diff-SERF AMs allow for operation in vapors with the conditions of high optical densities, insensitivity to AC-Stark shift noise, and insensitivity to magnetic fields from a secondary direction. Finally, a reduction of AC-Stark effects will be of considerable advantage in constructing  gradiometers with good common-mode rejection.

This work was supported by the NIH Eunice Kennedy
Shriver National Institute of Child Health \& Human
Development, \#R01HD057965. The authors are solely
responsible for the content.

\end{document}